# The Challenges of Balancing AI Compliance and Technological Innovations in Critical Sectors: A Systematic Literature Review

Ayush Enkhtaivan
Claremont Graduate University
ayush.enkhtaivan@cgu.edu

Chinazunwa Uwaoma
Claremont Graduate University
chinazunwa.uwaoma@cgu.edu

## Abstract

*The rapid integration of artificial intelligence (AI) into critical infrastructure including healthcare, finance, energy, and defense, offers transformative benefits but also conflicts with evolving regulatory and governance frameworks. This paper presents a systematic literature review (SLR) to examine the challenges of balancing AI compliance and technological innovation across critical infrastructure sectors. The review follows established SLR guidelines to extract and synthesize insights from peer-reviewed articles, report, and institutional sources published between 2020–2025. The study identifies three interrelated challenges: fragmented regulations, excessive compliance burdens for smaller to medium enterprises (SMEs), and misaligned governance models. To address these challenges, the study highlights practical governance strategies, including risk-tiered regulation, compliance-by-design, and explainable AI, to support scalable and trustworthy AI deployment in critical sectors. Key contributions include a concise mapping of core AI-governance challenges and a conceptual diagram illustrating their overlap, as well as actionable strategies for policymakers and practitioners to harmonize oversight with innovation.*

## 1. Introduction

The AI revolution has caused dramatic changes in many sectors including the critical infrastructure. Today, we are actively embedding AI into the fabric of our critical infrastructure, entrusting it with decisions in healthcare, finance, manufacturing, and energy that directly affect human lives. The promise is profound: AI-powered diagnostic tools can accelerate disease detection and personalize treatments (Topol, 2019), algorithmic systems can protect financial markets (Lam, 2025; Habbal et al., 2025; Kuppan et al., 2024), and smart grids can optimize energy for a sustainable future (DHS, 2024). These capabilities augment operational efficiency and predictive power (Wang & Wu, 2024).

Nonetheless, this integration creates a high-stakes sociotechnical environment. For every life-saving algorithm or secured transaction, there is a commensurate risk of failure. However, the challenge is not merely technical but one of stewardship. One of the most critical governance problems of our time is ensuring that these complex, autonomous systems are safe, ethically aligned, and worthy of the public's trust. (Raji et al., 2020).

The core problem is a dangerous gap between the rapid pace of AI innovation and the deliberate pace of human governance. AI capabilities are evolving faster than our ability to institutionalize the legal, ethical, and safety frameworks required to manage them (Anderljung et al., 2023). While foundational initiatives like the EU AI Act and the U.S. DHS AI Safety Guidelines are emerging, they often remain a step behind the technology, thus, creating policies for yesterday's risks (European Parliament, 2023; DHS, 2024).

This disconnects forces practitioners - the engineers, data scientists, and leaders building these systems into an untenable position. They are caught between two conflicting mandates: innovate aggressively to solve urgent problems yet guarantee alignment with a regulatory landscape that is still under construction (Tartaro et al., 2023; Morrison Foerster, 2025). This misalignment manifests in two major ways: (1) *Paralysis by Over-regulation; where* rigid, slow-moving compliance burdens stifle the deployment of beneficial technologies, leaving life-saving innovations on the shelf in fields like healthcare and emergency response (Risk Management Magazine, 2025; Chen et al., 2023). (2) *Peril from Under-regulation;* where unvalidated or biased algorithms are deployed in high-impact environments, exposing populations to discriminatory outcomes, security breaches, or catastrophic safety failures (Wang & Wu, 2024; Raji et al., 2020).

There is also a strategic impediment that threatens the very scalability and social license of AI. Organizations are struggling to navigate a compliance corridor that is simultaneously restrictive and ill-defined (Wang & Wu, 2024). In mission-critical sectors, the inability to bridge the innovation-governance gap erodes public trust, invites blunt regulatory overreach, and ultimately stalls progress. As the stewards of these powerful technologies, we require a new paradigm for governance; one that is adaptive, risk-informed, and fundamentally human-centered.

Thus, the purpose of this study is to examine how organizations in critical infrastructure sectors manage the tension between compliance and innovation in AI deployment, and to identify actionable strategies for improving AI governance in practice**.** The specific objectives of this study are as follows:

- Assess the extent to which current AI governance frameworks address compliance, ethics, and innovation imperatives
- Identify implementation gaps and sector-specific constraints in regulated environments
- Synthesize themes from the literature to map common governance challenges across critical infrastructure sectors
- Identify integrated actionable strategies that enable both regulatory alignment and technological agility

To achieve these objectives, this study addresses the key research question: How do regulatory requirements influence the development, deployment, and innovation of AI technologies in these sectors?

The study investigates and synthesizes how AI governance frameworks vary across sectors and regions and identifies a thematic approach for balancing oversight with innovation.

The rest of the paper is structured as follows: Section 2 provides a background to the study. Section 3 describes the methodology used to conduct the SLR. Section 4 presents the results and the discussion, while Section 5 provides the conclusion and future work.

## 2. Background

### 2.1 AI Governance and Compliance

AI governance encompasses the frameworks, protocols, and organizational structures that ensure AI systems are designed, deployed, and monitored in ways that are ethically sound, legally compliant, and technically robust. As Batool et al. (2025) highlighted in their study, these governance mechanisms are not confined to one level; they stretch across everything from internal company policies right up to national laws and international agreements. Their research underscored that accountability, transparency, and a focus on the entire AI lifecycle are the absolute bedrock of truly responsible AI governance.

Despite increasing attention to AI ethics, practical and enforceable governance mechanisms remain underdeveloped in many sectors. Batool, Zowghi, and Bano (2023) argue that while many discussions around AI ethics stay in the realm of theory, there is a shortage of practical, enforceable models, especially ones that directly tackle risks specific to different industries. The authors recommended that practitioners embed governance directly into every phase of the AI lifecycle, covering everything from the source of data, the design of algorithms and the deployment and importantly, the ongoing monitoring once AI is deployed.

To address this implementation gap between ethical intent and operational practice, the National Institute of Standards and Technology (NIST) developed the AI Risk Management Framework (AI RMF 1.0). The core message of this framework (Tabassi, 2023) is about creating AI systems we can truly rely on. The framework identifies the key characteristics of a trustworthy AI, which are reliability, security, fairness, accountability and explainability (Tabassi, 2023). Most importantly, it promotes a risk-based approach that encourages continuous evaluation and adaptation, rather than rigid compliance to fixed rules, making it particularly relevant in critical, rapidly evolving environments.

### 2.2. Innovation Pressure in Critical Sectors

In critical sectors like healthcare, defense, and finance, the pressure to adopt AI technologies is not just driven by competitive advantages, but by operational necessity. Nelson et al. (2023) provided examples that show how AI in predictive maintenance and automated quality checks has completely transformed productivity in the manufacturing sector. Also in the medical field, fascinating techniques like federated learning and explainable AI (XAI) are being proposed to boost both patient privacy and the transparency of AI systems, as noted in (Okolo, et al., 2022).

However, these same sectors are also under immense pressure from strict regulatory obligations. Wang and Wu (2024) highlighted the delicate balance between fostering innovation and maintaining adequate oversight. Excessively rigid regulatory regimes can restrain experimentation and delay deployment, while insufficient governance raises the risk of harmful or biased outcomes. Their findings call for harmonized global frameworks that support innovation while enabling clear and enforceable compliance pathways.

The Department of Homeland Security (DHS) provided a clear roadmap with the *"Roles and Responsibilities Framework for Artificial Intelligence in Critical Infrastructure."* This document provides guidance on the critical need to deploy AI safely and securely in essential areas like energy, transportation, and communications. The framework explicitly positions the responsibility on AI developers themselves (DHS, 2024), where they are advised to actively evaluate any potentially dangerous features in their products, to ensure that the AI tools align with human values and protect user privacy.

### 2.3. Risk-Based Regulation and Global Disparities

Global approaches to AI regulation vary significantly, reflecting divergent policy priorities and governance philosophies. For instance, Al-Maamari (2025) did a comparative study on AI risk classification systems across the EU, U.S., UK, and China. His findings highlight varying priorities; some leaning more towards innovation, others towards ethics, and still others emphasizing public safety. What emerges is that while the EU tends to favor a more precautionary approach, the U.S. seems to prefer innovation-led self-regulation. This difference creates a lot of complexities for global companies trying to operate across multiple jurisdictions.

To address the unique risks posed by advanced AI systems, Anderljung et al. (2023) introduced the concept of "Frontier AI Regulation," which specifically targets those high-risk AI models that could have systemic impacts. They suggest building up robust compliance infrastructure through practical steps like registration protocols, regular risk reporting, and independent third-party audits. The goal is to strike a sensible balance by enabling groundbreaking research while actively mitigating potential societal risks.

Ethical guidance from the European Parliament (2020) further highlights some key qualities for AI, especially when these systems start taking over human decision-making. This study emphasizes on "safety mindset" on AI development by prioritizing that AI needs to be safe, trustworthy, reliable, and maintain its integrity. The view also aligns with calls for anticipatory governance, where designers consider the downstream impacts of AI systems before deployment.

In parallel, the World Economic Forum (WEF) has weighed in with a white paper titled "Artificial Intelligence and Cybersecurity: Balancing Risks and Rewards." This document zeroes in on the growing intersection of AI and cybersecurity. It stresses that while we absolutely need to manage the cyber risks that come with AI, it is equally important to keep fostering innovation in this space (Creese & Jurgens, 2025).

The recent research work reviewed in the sections above all point to the need to create a balance in AI regulations and technological innovations to make most of the benefits in the adoption of AI in critical infrastructures. In what follows, we discuss the methods used in the SLR.

## 3. Methods

When we set out to conduct this systematic literature review (SLR), we closely followed the well-established methodology developed by Okoli (2015). We settled for Okoli (2015) SLR guide because it cross-references Levy & Ellis (2006) and Kitchenham & Charters (2007) guides which have been widely used in IS domain. Okoli's approach is well-suited for research in information systems and governance, as it lays out eight distinct steps we used for selecting, evaluating, and bringing together existing research. This detailed process ensures our review is transparent, reproducible, and academically rigorous every step of the way. These eight steps are categorized under four main stages: Planning, Selection, Extraction, and Reporting as shown in Figure 1.

### 3.1. Planning Stage

Step 1: Purpose of the Literature Review. The purpose of this review is to explore the complex challenges and trade-offs involved in striking a balance between AI compliance and technological innovation, especially within our critical infrastructure sectors like healthcare, finance, energy, and defense.

The aim is to consolidate what is already known, pinpoint any gaps in the current research, and ultimately provide valuable insights for both practitioners and policymakers on effective governance frameworks and best practices.

Step 2: Protocol and Training. To keep our review consistent, we first developed a review protocol. This document outlined exactly how we decide what research to include or exclude, our search strategy, and how we pull out the relevant data. We refined this protocol iteratively to make sure it perfectly aligned with our research goals.

### 3.2. Selection Stage

Step 3: Practical Screening. Next, we conducted an initial screening primarily by looking at titles and abstracts. Studies made the cut if they met the following criteria:

- Published in English between 2020 and 2025.
- Focused on AI systems specifically within regulated or critical sectors.
- Discussed regulatory compliance, risk governance, innovation policies, or AI deployment frameworks.

Step 4: Search Literature. Our search strategy involved a comprehensive look at academic and institutional sources published from 2020 to 2025. We cast a wide net, utilizing a combination of both academic databases and policy-focused repositories:

- Digital Libraries: IEEE Xplore, ScienceDirect, ProQuest, and ACM Digital Library.
- Policy and gray literature sources were: NIST, DHS, World Economic Forum, EU Parliament, AI.gov, Scrut.io, EXIN, Blue Prism, and the Ada Lovelace Institute.

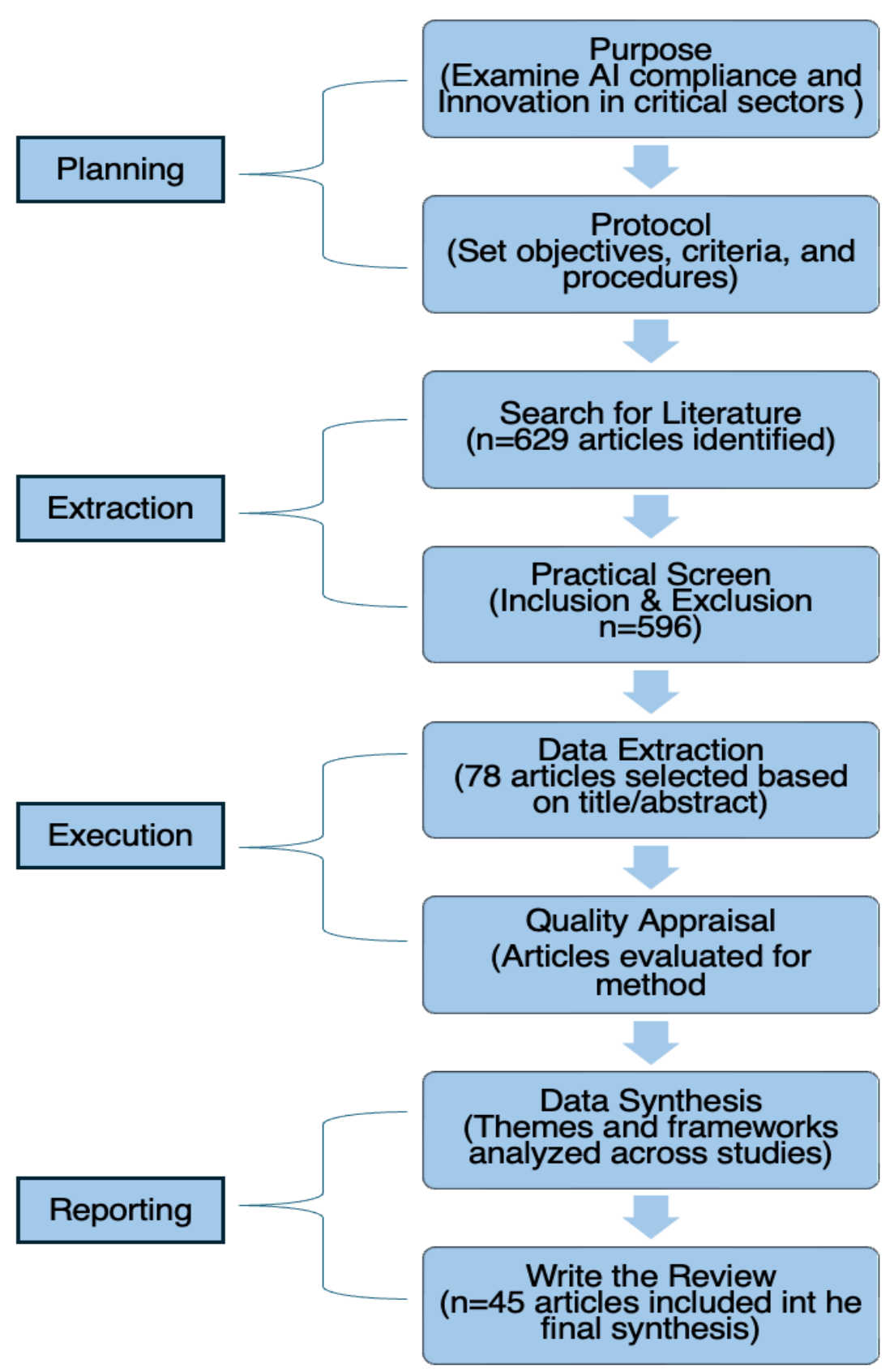

**Figure 1. Flow Diagram of the Selection Process**

**Keywords used:**

- "AI OR artificial intelligence compliance" AND "critical infrastructure"
- "AI OR artificial intelligence regulation" AND innovation
- "AI OR artificial intelligence governance" AND "regulation" AND "healthcare" OR "energy"
- "Responsible AI OR artificial intelligence" AND "innovation" AND "cybersecurity"
- "AI OR artificial intelligence risk management" AND "framework"

All the search results were then exported and de-duplicated using Zotero application.

### 3.3 Extraction Stage

Step 5: Data Extraction. We used a structured form to systematically extract key information from each selected study. This included:

- The author(s), year, title, and where it was published.
- Its specific sectoral focus. In the selection, we considered only healthcare, finance, energy, and defense sectors as these are consistently designated as critical in both the EU Directive and the US DHS taxonomy, and they also have explicit AI-relevant governance instruments.
- The type of compliance being discussed.
- Any challenges or benefits related to innovation.
- Details on governance frameworks, risk strategies, and policy recommendations.

Step 6: Quality Appraisal. After the initial screen, we performed a basic methodological appraisal. This helped us assess each source:

- Relevance to our specific research questions.
- Clarity of its arguments.
- Depth of its analysis (for example, whether it provided a solid conceptual framework or empirical basis).
- Publication type (prioritizing peer-reviewed articles but also including authoritative institutional reports).

Only high- and medium-quality sources were kept for further analysis. We made a specific point to include institutional white papers (like those from NIST and DHS) if they offered comprehensive and authoritative insights into governance frameworks.

### 3.4. Reporting

Step 7: Synthesize the Studies. With all the data extracted, we carefully went through the information we pulled from the articles collected. We examined each article by looking at its arguments, key findings, and specific contributions regarding AI compliance and innovation within critical infrastructure sectors.

This whole process involved several layers: first, we summarize the core insights from each source; then, we compared how different sectors were actually approaching regulatory compliance; and finally, we identified challenges that were shared across various industries.

We made sure to document the key findings, especially focusing on how regulatory demands were impacting the development and deployment of AI technologies, as well as any proposed solutions for navigating these constraints. Our synthesis highlights the burdens that complex governance requirements can impose, and the innovative practices organizations are already adopting to align compliance with technological advancement.

In addition to thematic synthesis, we mapped the 45 studies across four dimensions by: (i) country or region of origin, (ii) main entity of the study - unit of analysis (iii) research paradigm - qualitative, quantitative, mixed), and (iv) theoretical foundations. Guided by existing theoretical and governance frameworks (EU AI Act, NIST RMF), this inductive approach did not only allow us to identify regional and methodological gaps in the literature but also helped in identifying recurring concepts across studies which led to the derivation of the three overarching challenges broadly discussed in Section 4.

Step 8: Writing the Review. We present our findings by putting together all our findings in a structured and coherent way by including a detailed description of our entire review process, a clear presentation of our search results and the studies we ultimately selected, and a comprehensive discussion of both the challenges we observed and the responses that have been proposed in the literature.

## 4. Results and Discussions

In this section, we present the findings of this research. The final review is based on the 45 articles that was extracted from IEEE Xplore, ScienceDirect, ProQuest, and ACM. The search for the articles was carried out during the months of April and May 2025. A total number of 629 articles were retrieved based on the keywords. However, 78 relevant articles were selected after reviewing titles and using the delimitation criteria – Document Type: Conference and Journal Articles, Publication Year: 2020-2025, Language: English. Then we narrowed it down to total of 45 unique articles by eliminating duplicates as shown in Table 1. The distribution of the selected articles based on the digital repositories and the publication year is illustrated in Figure 2.

**Table 1. List of sources, keywords, and hits obtained**

| Digital Library | Search Keywords | # of Hits | Selected Articles |
|---|---|---|---|
| IEEE Xplore | artificial intelligence OR AI compliance AND critical infrastructure<br>AI OR artificial intelligence regulation AND innovation | 134 | 12 |
| ScienceDirect | AI OR artificial intelligence governance AND technological innovation<br>responsible artificial intelligence OR AI AND critical sectors | 220 | 35 |
| ProQuest | Artificial Intelligence OR AI AND compliance AND healthcare OR energy<br>AI OR artificial intelligence ethics AND regulatory frameworks | 178 | 23 |
| ACM Digital Library | AI OR artificial intelligence policy AND digital transformation<br>AI OR artificial intelligence regulation AND implementation challenges | 97 | 8 |
| Total | | **629** | **78** |
| Eliminated Duplicates | | | 33 |
| Total articles selected from the search | | | **45** |

Figure 3 on the other hand, shows the distribution of the studies across different regions and sectors. Whereas AI ethics and compliance is a global concern, Europe and American continents are in the forefront of research bordering on these issues. It is also evident from the graph that studies cut across many sectors with energy and healthcare receiving most of the attention.

We further categorized the reviewed studies in three key dimensions to highlight the methodological coverage as illustrated in Table 2.

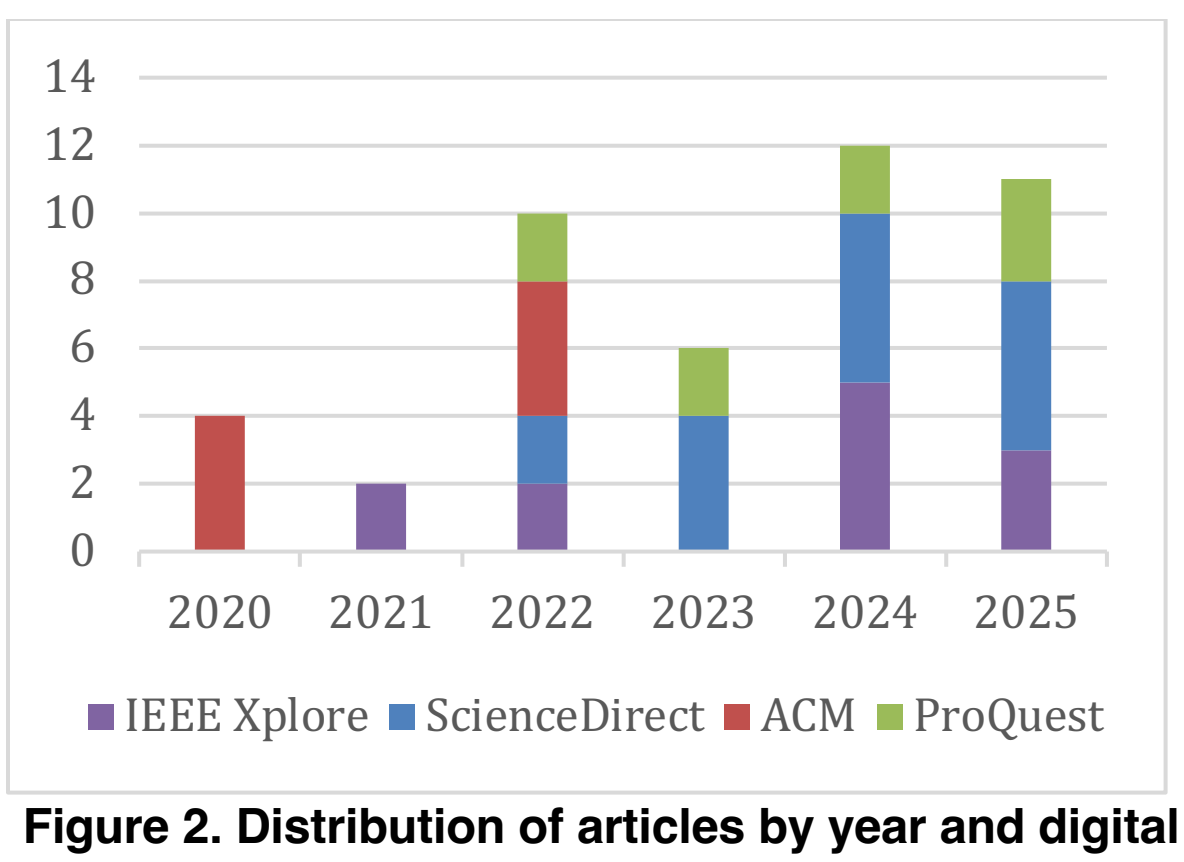

**Figure 2. Distribution of articles by year and digital library**

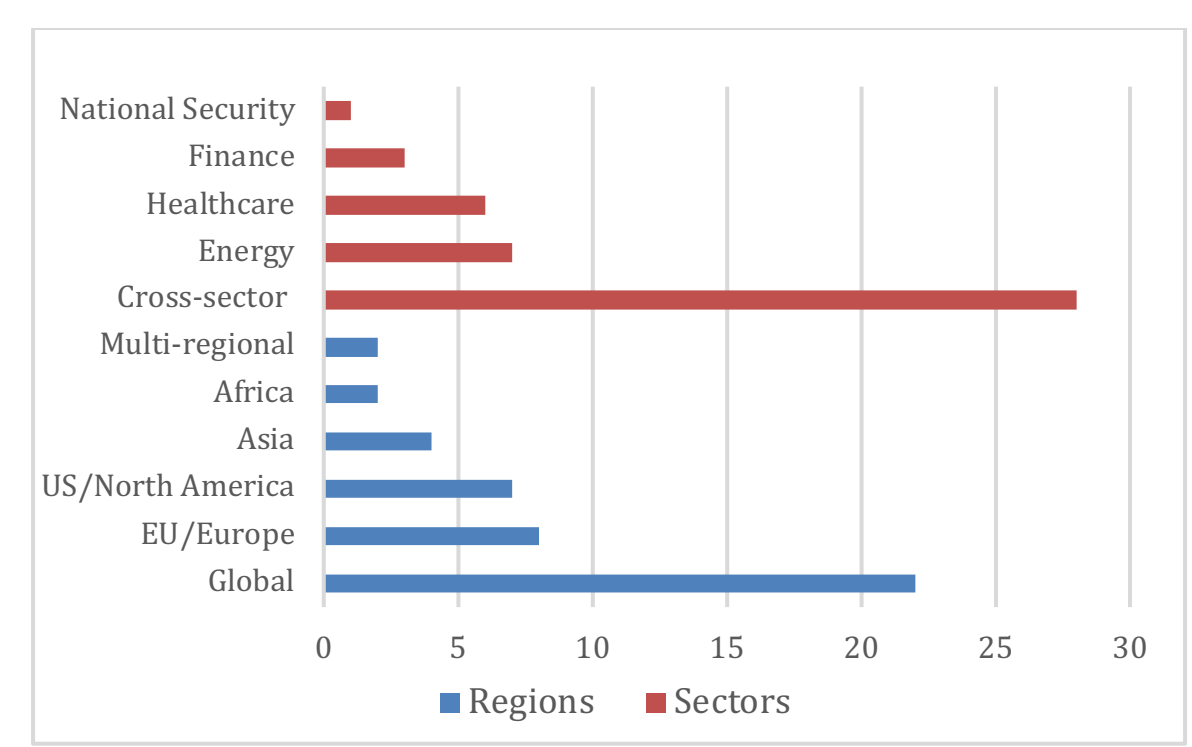

**Figure 3. Distribution of articles by region and sector**

**Table 2. Categorization of articles based on observed entity and methodological leaning**

| Unit of Analysis | Research Paradigm | Theoretical Foundations | References |
|---|---|---|---|
| Individual | Not observed. | N/A | |
| Organizational | Qualitative case / comparative studies; engineering evaluations; occasional mixed methods. | STS & workflow fit; clinical safety/XAI; socio-materiality; TTF; DeLone–McLean IS success; dynamic capabilities; model risk management. | Kannelønning, 2024; Bach et al., 2024; Sendak et al., 2024; Okolo et al., 2022; Laplante & Amaba, 2021; Mitchell et al., 2021; Malakar et al., 2025; Roshanpour et al., 2025; Lam, 2025; Kuppan et al., 2024. |
| National | Conceptual/policy analyses; institutional/grey frameworks; SLRs on regimes and standards. | Risk governance & standards; institutional theory; comparative governance/policy design; ethics/principlism. | Tabassi, 2023; Bird et al., 2020; Framework to Advance AI Governance in National Security (White House, 2024); Al-Maamari, 2025; Anderljung et al., 2023; Wang & Wu, 2024; Creese & Jurgens (WEF), 2025; Tartaro et al., 2023. |

## 4.1 Overview of Key Findings

This section presents the results and findings of the study. Our review uncovered several commonly observed challenges and solutions concerning AI compliance and innovation across different sectors. The study specifically identified three core, interrelated challenges that hinder responsible AI innovation in critical infrastructure sectors. These are: **Fragmented and Reactive Regulatory Landscapes**, **Innovation Impediments for SMEs**, and **Misalignment of Governance Models**.

These challenges do not operate in isolation; rather, they compound one another. This interrelationship is shown in the Venn diagram in Figure 4. Central to these Challenges are *Innovation Stagnation & Risk Exposure* where organizations especially smaller ones may delay or abandon AI development or proceed without adequate safeguards; thus, increasing both innovation inertia and systemic risk. At the intersection of Regulatory Fragmentation and Governance Misalignment are *Compliance Confusion & Delayed Implementation*, where organizations struggle to interpret conflicting or evolving standards, which leads to delays and inconsistent risk mitigation efforts. The interrelationship between Governance Misalignment and SME Innovation Barriers is seen as *High Compliance Burden,* where rigid governance structures often require documentation and oversight mechanisms that disproportionately affect SMEs with limited capacity. Key issues that cut across Regulatory Fragmentation and SME Barriers are *Unclear Guidelines & Overhead*, where SMEs frequently face a lack of accessible, harmonized compliance pathways, leading to increased costs and entry barriers.

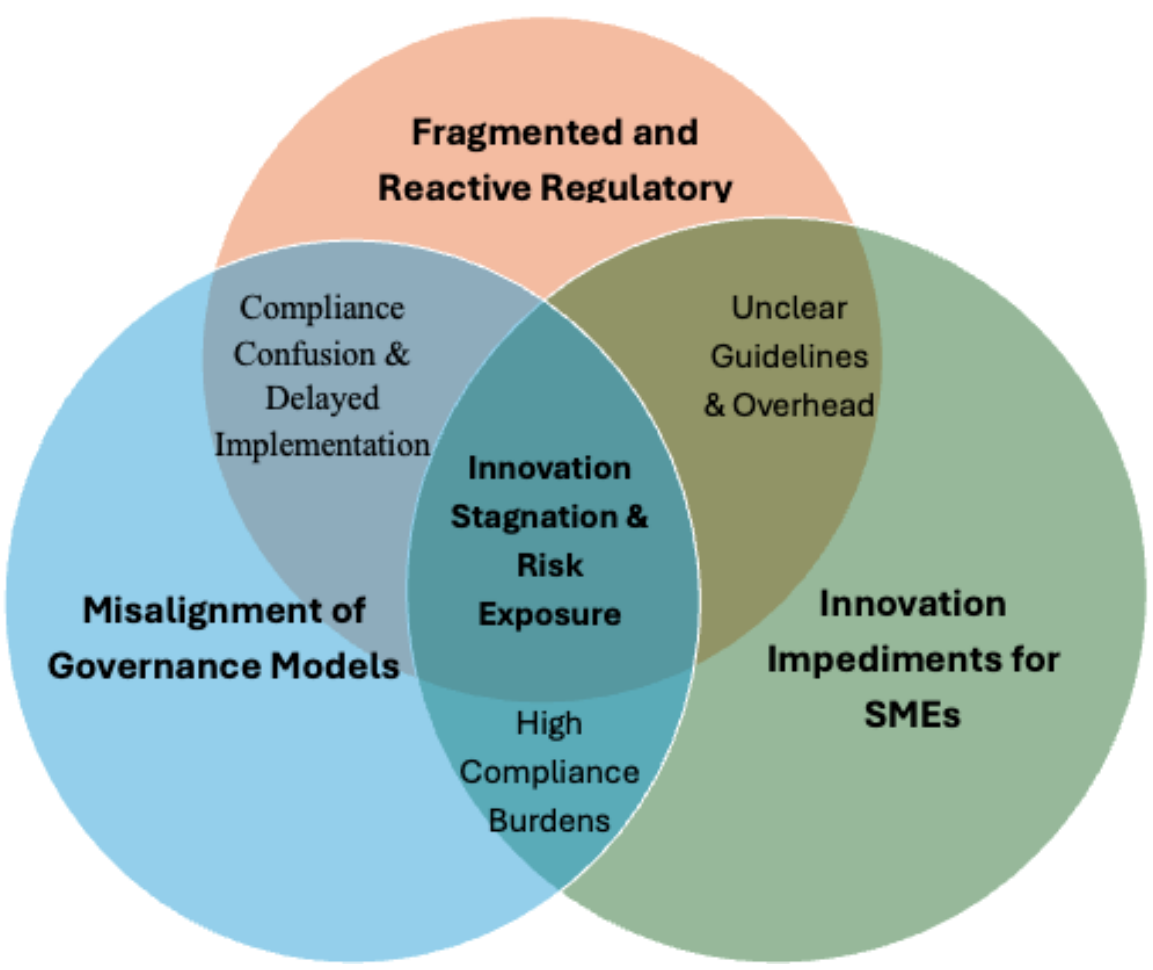

**Figure 4. Interrelationship between the core challenges**

The following subsections highlight each of these challenges and the suggested solutions based on the reviewed literature.

**4.1.1 Fragmented and Reactive Regulatory Landscapes.** Regulatory frameworks such as the EU AI Act and various national guidelines often lag behind technological advances, producing a patchwork of overlapping requirements that are difficult for organizations to interpret and implement. Many of the studies we looked at emphasized just how complex and fragmented our regulatory environments are, especially in high-stakes sectors like healthcare and energy. It is a significant challenge for organizations to navigate what are often overlapping, rapidly evolving, or very sector-specific compliance obligations. This often leads to frustrating delays in both AI development and its actual deployment, as highlighted by researchers in (Batool et al., 2025; Wang & Wu, 2024).

To help tackle these complexities, we have seen the introduction of more formal governance models. For instance, the NIST AI Risk Management Framework (RMF 1.0) offers structured guidance on how to design AI systems that align with key principles like safety, transparency, and fairness (TABASSI, 2023). Similarly, the U.S. Department of Homeland Security (DHS) has put forward its own framework for safely deploying AI across critical infrastructure sectors, with a strong focus on operational resilience and mitigating threats (DHS, 2024). These frameworks clearly signal a move towards proactive, risk-informed governance models, which is a positive step.

**4.1.2. Innovation Impediments for SMEs.** Smaller organizations, in particular, face prohibitive compliance costs and documentation burdens, which stifle experimentation and delay time-to-market for beneficial AI applications in areas like healthcare diagnostics and smart-grid management. Several articles pointed out a significant concern: unclear or overly rigid compliance frameworks can actually put a real damper on innovation. This is particularly true for SMEs that often just do not have the financial or technical muscle to meet complex legal demands (Thakkar, 2024; O'Brien et al., 2024). We frequently saw common barriers cited, such as high compliance costs, extensive audit and documentation requirements, and even the sheer fear of legal penalties. These factors can seriously discourage early-stage experimentation or the public deployment of exciting new AI models.

**4.1.3 Misalignment of Governance Model.** Rigid, one-size-fits-all compliance mandates frequently conflict with the agility needed for responsible innovation, leading either to "paralysis by over-regulation" or too risky "under-regulation" outcomes. In heavily regulated sectors like healthcare and energy, these constraints translate into tangible setbacks like delayed product launches, prototypes that get abandoned, or an excessive reliance on legacy systems. While compliant, the systems do not fully leverage the incredible capabilities AI offers (Creese & Jurgens, 2025). So, it is clear that compliance is often perceived as a trade-off against the speed and flexibility needed for true innovation.

**Table 3. Different patterns of AI regulations in different sectors**

| Sectors | Focus | Regulatory Patterns |
|---|---|---|
| Healthcare | Data privacy, transparency, explainability. | Strong adherence to data standards (e.g., HIPAA)<br>Emphasis on explainable AI to support clinical trust and ethical practice (Veenendaal, n.d.) |
| Finance | Algorithmic accountability, bias prevention, and fraud detection. | Development of auditable and fair AI systems<br>Prevention of Discriminatory outcomes<br>Need for sector-specific regulatory oversight |
| Energy | National security, cybersecurity risks, and resilient. | Emphasis on robustness and trustworthiness<br>Alignment with critical infrastructure protection polices<br>AI regulation linked with cyber-physical system safety (Hamayun, 2024) |

**Table 4. Distribution of the observed themes in different sectors**

| Theme | Sector | # of Studies | References |
|---|---|---|---|
| Fragmentation | Healthcare | 2 | Imaduddin et al., 2023; Nankya et al., 2024 |
| | Finance | 2 | Lam, 2025; Wang & Wu, 2024 |
| | Energy | 3 | Yigit et al., 2024; Adewusi et al., 2024; Creese & Jurgens (WEF), 2025 |
| | Defense / National Security | 3 | White House (Framework to Advance AI Governance in National Security), 2024; Adewusi et al., 2024; Anderljung et al., 2023 |
| | Cross-sector | 10 | Tabassi, 2023; Al-Maamari, 2025; Wang & Wu, 2024; Bird et al., 2020; Birkstedt et al., 2023; Bilan et al., 2022; Chen et al., 2023; Tartaro et al., 2023; Creese & Jurgens (WEF), 2025 |
| Misalignment | Healthcare | 4 | Sendak et al., 2024; Kannelønning, 2024; Mohsin Khan et al., 2025; Nwobodo-Anyadiegwu et al., 2022 |
| | Finance | 2 | Kuppan et al., 2024; Lam, 2025 |
| | Energy | 4 | Laplante & Amaba, 2021; Al-Hawawreh et al., 2024; Mitchell et al., 2021; Ahmad Hamdan et al., 2024 |
| | Cross-sector | 10 | Raji et al., 2020; Papagiannidis et al., 2025; Batool et al., 2025; Bach et al., 2024; Mariani et al., 2023; Malakar et al., 2025; Nelson et al., 2023; Roshanpour et al., 2025; Martini et al., 2024; Zheng et al., 2022 |
| SME | Finance | 4 | Habbal et al., 2025; Lam, 2025; Akheel, 2023; Patel, 2024 |
| | Cross-sector | 4 | Mosli, 2024; Akheel, 2023; Patel, 2024; Pattanayak, 2021 |

To overcome these challenges, several studies have proposed various governance models and best practices specifically designed to support innovation *without* compromising safety or accountability. One promising approach is adopting "compliance-by-design" methodologies. This essentially means embedding legal and ethical considerations right from the very beginning of the AI development lifecycle (Thales Group, 2025). The big advantage here is that it minimizes costly rework down the line and can significantly boost regulators confidence during system audits.

Other smart strategies include using explainable AI (XAI) to boost transparency and ensure humans can maintain oversight (Okolo et al., 2022). We also saw recommendations for risk-tiered governance frameworks, which intelligently adjust regulatory requirements based on how the AI system is intended to be used and its potential impact. The overarching goal of these practices is to cultivate a governance environment that is both scalable and adaptable, ultimately enabling responsible innovation even within those complex regulatory landscapes.

### 4.2 Sector-Specific Observations

Table 3 shows distinct patterns of AI regulations in different sectors as observed from the reviewed studies. Relatedly, the distribution of the themes across different sectors is highlighted in Table 4, which shows that the key challenges run across sectors but more pronounced in health and energy sectors.

## 5. Conclusion and Future Work

The SLR of scholarly and institutional sources conducted in this study highlighted the persistent tension between regulatory compliance and technological innovation in AI deployments across critical infrastructure sectors. Our findings reveal three core challenges: Fragmented and Reactive Regulatory Landscapes, Innovation Impediments for SMEs and, Misalignment of Governance Models Conversely, our review also identified promising practices that practitioners and policymakers can adopt to bridge this innovation-governance gap which include: Risk-Tiered Governance - focusing resources on high-risk applications while lightening the load on lower-risk use cases; Compliance-by-Design - embedding legal, ethical, and safety considerations early in the AI lifecycle to reduce costly retrofits and enhance regulator confidence; and Explainable and Monitored AI (XAI) - leveraging transparency tools and continuous monitoring to maintain human oversight and accountability without unduly constraining innovation. By synthesizing these insights, this review offers a roadmap for critical-sector organizations to implement AI solutions that are both compliant and cutting-edge, ultimately fostering public trust and enabling scalable, responsible innovation.

## 5.1 Study Limitations

It is important to recognize that research in AI regulations and compliance is still relatively new and on-going; hence, the need to note some of the limitations we had in conducting the study. The review was confined to publications between 2020 and 2025 and included only English-language sources, potentially omitting relevant earlier works or non-English studies that could offer additional insights. Also, our search drew exclusively from four digital libraries and selected institutional websites. This focus may have overlooked pertinent research in other repositories or in less formal venues, such as industry white papers or conference proceedings not indexed in these platforms. By synthesizing findings across healthcare, finance, energy, and defense, the review emphasizes high-level, cross-sector themes. This broad perspective may understate unique, sector-specific governance challenges and best practices. These limitations hint on future work to broaden the linguistic and database coverage and periodically update the review to reflect the real-world vertical impact of the identified challenges and the solutions in specific sectors.

## 5.2 Future Work

To extend this study, we will explore several avenues for future research. These include conducting sector-specific case studies to measure the real-world effectiveness of risk-tiered and compliance-by-design approaches, to quantify both innovation gains and compliance outcomes. We will also look at creating methodologies to assess the economic trade-offs of different compliance strategies, that will enable organizations especially SMEs, to make data-driven decisions about investment in governance versus innovation. Another area we will work on is to track how recent regulatory developments and the implementations influence innovation pipelines over time, informing adaptive policy adjustments that better balance safety and agility. Pursuing these research directions will no doubt, deepen our understanding of how to craft AI governance mechanisms that not only safeguard public interests but also unleash the full potential of AI-driven innovation in critical sectors.

# 6. References

Adebunmi Okechukwu Adewusi, Ugochukwu Ikechukwu Okoli, Temidayo Olorunsogo, Ejuma Adaga, Donald Obinna Daraojimba, & Ogugua Chimezie Obi. (2024). Artificial intelligence in cybersecurity: Protecting national infrastructure: A USA review. World Journal of Advanced Research and Reviews, 21(1), 2263–2275. https://doi.org/10.30574/wjarr.2024.21.1.0313

Ahmad Hamdan, Kenneth Ifeanyi Ibekwe, Valentine Ikenna Ilojianya, Sedat Sonko, & Emmanuel Augustine Etukudoh. (2024). AI in renewable energy: A review of predictive maintenance and energy optimization. International Journal of Science and Research Archive, 11(1), 718–729. https://doi.org/10.30574/ijsra.2024.11.1.0112

AI Compliance: What It Is and Why You Should Care [2024 update]. (2024). EXIN. https://www.exin.com/article/ai-compliance-what-it-is-and-why-you-should-care/

Akheel, S. A. (2023). AI Governance and Compliance in Regulated Industries. Journal of Artificial Intelligence & Cloud Computing, 2(4), 1–9. https://doi.org/10.47363/JAICC/2023(2)E217

Al-Hawawreh, M., Baig, Z., & Zeadally, S. (2024). AI for Critical Infrastructure Security: Concepts, Challenges, and Future Directions. IEEE Internet of Things Magazine, 7(4), 136–142. https://doi.org/10.1109/IOTM.001.2300181

Al-Maamari, A. (2025). Between Innovation and Oversight: A Cross-Regional Study of AI Risk Management Frameworks in the EU, U.S., UK, and China (arXiv:2503.05773). arXiv. https://doi.org/10.48550/arXiv.2503.05773

Anderljung, M., Barnhart, J., Korinek, A., Leung, J., O'Keefe, C., Whittlestone, J., Avin, S., Brundage, M., Bullock, J., Cass-Beggs, D., Chang, B., Collins, T., Fist, T., Hadfield, G., Hayes, A., Ho, L., Hooker, S., Horvitz, E., Kolt, N., … Wolf, K. (2023). Frontier AI Regulation: Managing Emerging Risks to Public Safety (arXiv:2307.03718). arXiv. https://doi.org/10.48550/arXiv.2307.03718

Bach, T. A., Kristiansen, J. K., Babic, A., & Jacovi, A. (2024). Unpacking Human-AI Interaction in Safety-Critical Industries: A Systematic Literature Review. IEEE Access, 12, 106385–106414. https://doi.org/10.1109/ACCESS.2024.3437190

Batool, A., Zowghi, D., & Bano, M. (2023). Responsible AI Governance: A Systematic Literature Review (arXiv:2401.10896). arXiv. https://doi.org/10.48550/arXiv.2401.10896

Batool, A., Zowghi, D., & Bano, M. (2025). AI governance: A systematic literature review. AI and Ethics, 5(3), 3265–3279. https://doi.org/10.1007/s43681-024-00653-w

Bilan, S., Šuleř, P., Skrynnyk, O., Krajňáková, E., & Vasilyeva, T. (2022). Systematic bibliometric review of artificial intelligence technology in organizational management, development, change and culture. Verslas : Teorija Ir Praktika, 23(1), 1–13. https://doi.org/10.3846/btp.2022.13204

Bird, E., Skelly, J. F., Jenner, N., Larbey, R., Weitkamp, E., & Winfield, A. (2020). The ethics of artificial intelligence: Issues and initiatives. Publications Office. https://data.europa.eu/doi/10.2861/6644

Birkstedt, T., Minkkinen, M., Tandon, A., & Mäntymäki, M. (2023). AI governance: Themes, knowledge gaps and

future agendas. Internet Research, 33(7), 133–167. https://doi.org/10.1108/INTR-01-2022-0042
Chen, C., Fu, J., & Lyu, L. (2023). A Pathway Towards Responsible AI Generated Content (arXiv:2303.01325). arXiv. https://doi.org/10.48550/arXiv.2303.01325
Creese, S., & Jurgens, J. (2025). WEF_Artificial_Intelligence_and_Cybersecurity_Balancing_Risks_and_Rewards_2025.pdf. World Economic Forum. https://reports.weforum.org/docs/WEF_Artificial_Intelligence_and_Cybersecurity_Balancing_Risks_and_Rewards_2025.pdf
Emerging Threats to Critical Infrastructure: AI Driven Cybersecurity Trends for 2025 | Capitol Technology University. (2025). Capital Technology Institute. https://www.captechu.edu/blog/ai-driven-cybersecurity-trends-2025
EU AI Act: First regulation on artificial intelligence. (2023, August 6). Topics | European Parliament. https://www.europarl.europa.eu/topics/en/article/20230601STO93804/eu-ai-act-first-regulation-on-artificial-intelligence
Framework to Advance AI Governance and Risk Management in National Security. (2024). White House.
Groundbreaking Framework for the Safe and Secure Deployment of AI in Critical Infrastructure Unveiled by Department of Homeland Security | Homeland Security. (2024). Department of Homaland Security. https://www.dhs.gov/archive/news/2024/11/14/groundbreaking-framework-safe-and-secure-deployment-ai-critical-infrastructure
Habbal, R., Slimani, I., & El Farissi, I. (2025). Artificial Intelligence in Financial Services: Applications, Challenges, and Research Directions. 2025 5th International Conference on Innovative Research in Applied Science, Engineering and Technology (IRASET), 1–6. https://doi.org/10.1109/IRASET64571.2025.11008019
Hamayun, M. (2024, September 25). AI: The New Frontier in Safeguarding Critical Infrastructure. Check Point Blog. https://blog.checkpoint.com/artificial-intelligence/ai-the-new-frontier-in-safeguarding-critical-infrastructure/
Imaduddin, A. M., Spooner, B., Isherwood, J., Lane, M., Emma, O., & Dennison, A. (2023). A Systematic Review of the Barriers to the Implementation of Artificial Intelligence in Healthcare. Cureus, 15(10). https://doi.org/10.7759/cureus.46454
Kannelønning, M. S. (2024). Contesting futures of Artificial Intelligence (AI) in healthcare: Formal expectations meet informal anticipations. Technology Analysis & Strategic Management, 36(11), 3845–3856. https://doi.org/10.1080/09537325.2023.2226243
Kitchenham, B., & Charters, S. (2007). Systematic literature reviews in software engineering – A systematic literature review. Information and Software Technology, 51(1), 7–15. https://doi.org/10.1016/j.infsof.2008.09.009
Knisley, J. (2025). Risk Management Magazine—Balancing Innovation and Compliance When Implementing AI. Magazine. https://www.rmmagazine.com/articles/article/2025/04/30/balancing-innovation-and-compliance-when-implementing-ai
Kuppan, K., Bhaskar Acharya, D., & B, D. (2024). Foundational AI in Insurance and Real Estate: A Survey of Applications, Challenges, and Future Directions. IEEE Access, 12, 181282–181302. https://doi.org/10.1109/ACCESS.2024.3509918
Lam, A. Y. S. (2025). Artificial Intelligence Applications in Financial Technology. Journal of Theoretical and Applied Electronic Commerce Research, 20(1), Article 1. https://doi.org/10.3390/jtaer20010029
Laplante, P., & Amaba, B. (2021). Artificial Intelligence in Critical Infrastructure Systems. Computer, 54(10), 14–24. https://doi.org/10.1109/MC.2021.3055892
Levy, Y., & J. Ellis, T. (2006). A Systems Approach to Conduct an Effective Literature Review in Support of Information Systems Research. Informing Science: The International Journal of an Emerging Transdiscipline, 9, 181–212. https://doi.org/10.28945/479
Malakar, P., Khan, S. A., Gunasekaran, A., & Mubarik, M. S. (2025). Digital Technologies for Social Supply Chain Sustainability: An Empirical Analysis Through the Lens of Dynamic Capabilities and Complexity Theory. IEEE Transactions on Engineering Management, 72, 664–675. https://doi.org/10.1109/TEM.2024.3524896
Mariani, M. M., Machado, I., Magrelli, V., & Dwivedi, Y. K. (2023). Artificial intelligence in innovation research: A systematic review, conceptual framework, and future research directions. Technovation, 122, 102623. https://doi.org/10.1016/j.technovation.2022.102623
Martini, B., Bellisario, D., & Coletti, P. (2024). Human-Centered and Sustainable Artificial Intelligence in Industry 5.0: Challenges and Perspectives. Sustainability, 16(13), 5448. https://doi.org/10.3390/su16135448
Mitchell, D., Blanche, J., Zaki, O., Roe, J., Kong, L., Harper, S., Robu, V., Lim, T., & Flynn, D. (2021). Symbiotic System of Systems Design for Safe and Resilient Autonomous Robotics in Offshore Wind Farms. IEEE Access, 9, 141421–141452. https://doi.org/10.1109/ACCESS.2021.3117727
Mohsin Khan, M., Shah, N., Shaikh, N., Thabet, A., alrabayah, T., & Belkhair, S. (2025). Towards secure and trusted AI in healthcare: A systematic review of emerging innovations and ethical challenges. International Journal of Medical Informatics, 195, 105780. https://doi.org/10.1016/j.ijmedinf.2024.105780
Mosli, R. (2024). Optimal Job Role Allocation for Compliance With NCA-ECC Controls Using the Saudi Cybersecurity Workforce Framework. IEEE Access, 12, 128235–128245. https://doi.org/10.1109/ACCESS.2024.3457025
Mühlroth, C., & Grottke, M. (2022). Artificial Intelligence in Innovation: How to Spot Emerging Trends and Technologies. IEEE Transactions on Engineering Management, 69(2), 493–510. https://doi.org/10.1109/TEM.2020.2989214
Nankya, M., Mugisa, A., Usman, Y., Upadhyay, A., & Chataut, R. (2024). Security and Privacy in E-Health Systems: A Review of AI and Machine Learning Techniques. IEEE Access, 12, 148796–148816. https://doi.org/10.1109/ACCESS.2024.3469215

Nelson, J. P., Biddle, J. B., & Shapira, P. (2023). Applications and Societal Implications of Artificial Intelligence in Manufacturing: A Systematic Review (arXiv:2308.02025). arXiv. https://doi.org/10.48550/arXiv.2308.02025

Nwobodo-Anyadiegwu, E. N., Ditend, M. N., & Lumbwe, A. K. (2022). The benefits and challenges of Implementing Smart hospital projects: A systematic Review. 2022 IEEE 28th International Conference on Engineering, Technology and Innovation (ICE/ITMC) & 31st International Association For Management of Technology (IAMOT) Joint Conference, 1–7. https://doi.org/10.1109/ICE/ITMC-IAMOT55089.2022.10033238

O'Brien, C., Rasdale, M., & Wong, D. (2024). The role of harmonised standards as tools for AI act compliance | DLA Piper. DLA Piper. https://www.dlapiper.com/en/insights/publications/2024/01/the-role-of-harmonised-standards-as-tools-for-ai-act-compliance

Okoli, C. (2015). A Guide to Conducting a Standalone Systematic Literature Review. Communications of the Association for Information Systems, 37. https://doi.org/10.17705/1CAIS.03743

Okolo, C. T., Dell, N., & Vashistha, A. (2022). Making AI Explainable in the Global South: A Systematic Review. ACM SIGCAS/SIGCHI Conference on Computing and Sustainable Societies (COMPASS), 439–452. https://doi.org/10.1145/3530190.3534802

Outlook on DHS Framework for AI in Critical Infrastructure. (2025). Morrison Foerster. https://www.mofo.com/resources/insights/null

Papagiannidis, E., Mikalef, P., & Conboy, K. (2025). Responsible artificial intelligence governance: A review and research framework. The Journal of Strategic Information Systems, 34(2), 101885. https://doi.org/10.1016/j.jsis.2024.101885

Patel, K. (2024). Ethical Reflections On Data-Centric AI: Balancing Benefits And Risks. SSRN Electronic Journal. https://doi.org/10.2139/ssrn.4993089

Pattanayak, S. K. (2021). Navigating Ethical Challenges in Business Consulting with Generative AI: Balancing Innovation and Responsibility. 10(2).

Pouget, H. (2023). What will the role of standards be in AI governance? Ada Lovelace Institute. https://www.adalovelaceinstitute.org/blog/role-of-standards-in-ai-governance/

Raji, I. D., Smart, A., White, R. N., Mitchell, M., Gebru, T., Hutchinson, B., Smith-Loud, J., Theron, D., & Barnes, P. (2020). Closing the AI Accountability Gap: Defining an End-to-End Framework for Internal Algorithmic Auditing (arXiv:2001.00973). arXiv. https://doi.org/10.48550/arXiv.2001.00973

Roshanpour, R., Parsanejad, M., & Saman Pishvaee, M. (2025). An AI-ISM Methodology for Structural Modeling of Sustainable Supply Chain Management Drivers. IEEE Access, 13, 76481–76496. https://doi.org/10.1109/ACCESS.2025.3559838

Sendak, M. P., Liu, V. X., Beecy, A., Vidal, D. E., Shaw, K., Lifson, M. A., Tobey, D., Valladares, A., Loufek, B., Mogri, M., & Balu, S. (2024). Strengthening the use of artificial intelligence within healthcare delivery organizations: Balancing regulatory compliance and patient safety. Journal of the American Medical Informatics Association, 31(7), 1622–1627. https://doi.org/10.1093/jamia/ocae119

Tabassi, E. (2023). Artificial Intelligence Risk Management Framework (AI RMF 1.0) (NIST AI 100-1; p. NIST AI 100-1). National Institute of Standards and Technology (U.S.). https://doi.org/10.6028/NIST.AI.100-1

Tartaro, A., Smith, A. L., & Shaw, P. (2023). Assessing the impact of regulations and standards on innovation in the field of AI (arXiv:2302.04110). arXiv. https://doi.org/10.48550/arXiv.2302.04110

Thakkar, M. (2023, December 15). AI compliance in 2025: Key regulations and strategies for business. Scrut Automation. https://www.scrut.io/post/ai-compliance

The challenges of using AI in critical systems | Thales Group. (2025, February 10). Thales Group. https://www.thalesgroup.com/en/worldwide/group/magazine/challenges-using-ai-critical-systems

The importance of AI standards: The key to innovation and compliance. (2025, March 31). Naaia. https://naaia.ai/standard-ai-compliance-innovation/

Topol, E. (2019). Deep medicine: How artificial intelligence can make healthcare human again. New York : Basic Books. https://catalog.nlm.nih.gov/discovery/fulldisplay/alma9917337203406676/01NLM_INST:01NLM_INST

Veenendaal, A. (2024). AI Compliance Best Practices. SS&C Blue Prism. https://www.blueprism.com/guides/ai/ai-compliance/

Wang, X., & Wu, Y. C. (2024). Balancing Innovation and Regulation in the Age of Generative Artificial Intelligence. Journal of Information Policy, 14, 385–416. https://doi.org/10.5325/jinfopoli.14.2024.0012

Yigit, Y., Ferrag, M. A., Sarker, I. H., Maglaras, L. A., Chrysoulas, C., Moradpoor, N., & Janicke, H. (2024). Critical Infrastructure Protection: Generative AI, Challenges, and Opportunities (arXiv:2405.04874). arXiv. https://doi.org/10.48550/arXiv.2405.04874

Zheng, Q., Tang, Y., Liu, Y., Liu, W., & Huang, Y. (2022). UX Research on Conversational Human-AI Interaction: A Literature Review of the ACM Digital Library. CHI Conference on Human Factors in Computing Systems, 1–24. https://doi.org/10.1145/3491102.3501855